%% file: root_V2.tex
\documentclass[journal,draftcls,onecolumn,12pt]{IEEEtran}

\usepackage{graphicx}
\usepackage{epstopdf}
\usepackage{amsmath}
\usepackage{epsfig}
\usepackage{amsthm}
\begin{document}
\title{A Theorem on Multi-Objective Optimization Approach for Bit Allocation of Scalable Coding}

\author{Wen-Liang Hwang  \\
Institute of Information Science, Academia Sinica, Taiwan }

\date{}
\maketitle \pagenumbering{arabic}

\begin{abstract}
In the current work, we have formulated the
optimal bit-allocation problem for a scalable codec of images or videos as a constrained
vector-valued optimization problem and demonstrated that there can be many optimal
solutions, called Pareto optimal points. In practice, the Pareto points are derived via the weighted sum scalarization approach. An important question which arises is whether all the Pareto optimal points can be derived using the scalarization approach?  The present paper provides a sufficient condition on the rate-distortion function of each resolution of a scalable codec to address the above question.
The result indicated that if the rate-distortion function of each resolution is strictly decreasing and convex and the Pareto points form a continuous curve, then all the optimal Pareto
points can be derived by using the scalarization
method.

\end{abstract}

\section{Introduction} \label{sec0}

Scalable coding (SC) involves producing from an image or a video (also called coding object) a single
bit-stream that meets user requirements of resolutions of the image or the video \cite{Sko2001,Ye14}. In SC,
the bit-stream is usually organized into subset bit-streams with various resolutions of the coding object.  The subset bit-streams
are generally correlated by prediction methods to enhance
coding efficiency \cite{Lu13,Zhongbo12}. 
The coding efficiency
can also be improved if the bit-allocation, which distributes an
available amount of bits to resolution, can be optimized
\cite{Sullivan13,Wie2003b,Kaaniche14}.

In scalable coding studies, the usual assumption is that the solution of the
bit-allocation optimization problem is either better or at least
no worse than any other alternative. However, this assumption is
only correct if all the users demand the same resolution and the coding object is compressed for that resolution.
For such a case, the optimization problem can be solved for that particular 
resolution, and all the users can receive the best service simultaneously 
from the coding system.
However, for SC, where a single bit-stream is designed to serve
many users with various demands of resolution, the performance
criteria for different resolutions clearly conflict. As a
result, the assumption that an optimum bit-stream can be achieved which 
would produce the best performance simultaneously for all the resolutions
is generally incorrect. Specifically, it is very unlikely that a bit-allocation which will optimize one resolution
will also optimize the other resolutions. 
The top subgraph of Figure \ref{Uncomparable} shows how two different
bit-allocations have been assigned to support three spatial resolutions, where the
left-most node supports quarter common intermediate format (QCIF), the left-most
and the middle nodes support CIF, and the three nodes all together support high definition (HD).
A same bit number has been assigned to the node of QCIF, therefore,
the distortion comparison for the two bit-allocations is on
CIF and HD. The bottom subgraph of Figure \ref{Uncomparable} shows
two distortions for CIF and HD with respect to the two
bit-allocations. On comparing the distortions of the two bit-allocations for both CIF and HD, it can be inferred that one bit-allocation is better for CIF, but worse
for HD, whereas the other is better for HD but not for CIF.
Figure \ref{Uncomparable} thus demonstrates that it is
not always possible for a bit-allocation procedure to generate a
bit-stream that can simultaneously achieve the best performance for all
the resolutions. Furthermore, since we may not determine that one resolution is more important than another,
the performance of any two bit-allocations is, in general, incomparable.

Since a scalable codec serves multiple resolutions simultaneously,
the performance of a bit-allocation cannot be measured with a single
objective function. Instead, it is a multi-objective (multi-criteria)
function, with a vector-valued objective, where each component of the
objective represents the performance of one resolution.
The definition of an optimal solution in a multi-objective problem
is referred to as Pareto optimality
\cite{Par,Ehr,Eic}. Intuitively, an optimal solution (called a Pareto point) reaches equilibrium in the objective vector space in the sense that any
improvement of a participant can only be obtained if there is deterioration 
of at least one other participant. Therefore, no movement
can raise the consensus by all the participating parties in the
equilibrium. Since the Pareto points cannot be ordered
and compared, it cannot be determined which point is better or worse
than the others. 

In SC, the participating parties are the resolutions,
and the objective space is the space of the performance of the
resolutions. 
The multi-criteria perspective is also supported by the weighted sum scalarization method  where the optimal
bit-allocation can be obtained by solving the weighted sum of the
distortions of resolutions:
\begin{equation}
\min_{\underline{b}\in \Omega} \sum_{i=0}^{N-1} w_i
g_i(\underline{b}), \label{liter}
\end{equation} where $\underline{b}\in \Omega$ is a feasible bit-allocation vector (or bit-allocation profile), and $w_i$
and $g_i$ are non-negative weight and distortion for the
resolution $i$, respectively. By varying the values of the weights
$w_i$, solving (\ref{liter}) yields different Pareto optimal points.
In general, the solutions of (\ref{liter}) form a subset of the
Pareto optimal points. Thus, the solutions of (\ref{liter}) cannot cover all the performance that a scalable coding method can achieve. Meanwhile, the Pareto optimal solution to the problem of scalable
coders is generally large, and if computational cost is a concern, the performance comparison of bit-allocation methods is usually set at a few Pareto points \cite{Ohm04,SchwarzMW07,SchwarzW07,Sul1998,Sullivan12,Chakareski13,Liu10,WangNash14}. The weight vector associated with (\ref{liter}) is either given or derived based on users' preference choice \cite{Peng12SVC,Peng12Wavelet}. 
In the literature of SV, solving the bit-allocation problem was mainly based on modelling the  rate-distortion (R-D) function $g_i(\underline{b})$ \cite{NJ84,Usevitch96,Mal1998,Woods99,Taubman00,schaar01,He2002}. The performance comparison, therefore, mainly comprised accuracy and efficiency of the rate-distortion models at some particular Pareto points. 

Since the Pareto points derived by using the weighted sum scalarization approach is widely used in SC to conduct performance comparison of bit-allocation methods and rate-distortion models, we  were motivated to derive the conditions under which the scalarization approach can cover all the Pareto points. The main result is shown in Theorem 2, which states that if the R-D function of each resolution is a strictly decreasing convex function and the Pareto points form a continuous curve, then all the Pareto points can be derived by using the scalarization approach. This result was derived based on formulating the SC's bit-allocation problem as a multi-objective optimization problem defined on a directed acyclic graph (DAG), representing the coding dependency of a codec. A discrete version of the theorem is also presented.

The main contributions of the current study are: 1) the bit-allocation problem for SC has been formulated as a multi-objective optimization problem. The optimal bit-allocation is a set of Pareto points; 2)  the rate-distortion (R-D) curve of each resolution of a SC has been characterized so that  all the (weakly) Pareto optimal points can be derived by using the weighted sum scalarization approach.

The rest of the paper is organized as follows. In Section
\ref{DAGsection}, we presented the prediction
structure of SC using a DAG. In Section \ref{secmop}, we formulated the optimal
bit-allocation problem of SC in a DAG and used the Pareto optimal points to
characterize the solutions of the problem. 
Section \ref{RDModel} contains the man results which characterize all the Pareto points from the R-D function of each resolution of a scalable coding method by using the scalarization approach.
Section \ref{con}
presents the concluding remarks. \\

\noindent{\textbf{Notations}}.

\noindent We have used underline to indicate a vector; for example, $x$ is
a scalar and $\underline{x}$ is a vector. Let $\underline{x} =
[x_i]^T$ and $\underline{y}=[y_i]^T$ be two vectors. The following
operations are defined based on the vector notation. \\
1. $\underline{x} \in R_{+}^N$ (the cone of nonnegative orthant in $R^N$) if $x_i \geq 0$ for all $i$. \\
2. $\underline{x} < \underline{y}$ if $x_i \leq y_i$ for all $i$,
and there is a $j$ such
that $x_j < y_j$. \\
3. $\underline{x} \leq \underline{y}$ if for all $i$ such that $x_i \leq y_i$. \\
4. $\underline{x} \ll \underline{y}$ if $x_i <  y_i$ for all $i$. \\
5. $\underline{x}^T$ is the transpose of the vector $\underline{x}$.

\section{Directed Graph Model for Data Dependency}
\label{DAGsection}

In SC, a coding object is usually divided into multiple coding segments.
The layers are the basic coding segments in SC that support spatial
and quality scalability in an image and spatial, temporal, and quality scalability in a video.
To remove the abundant redundancy existing between the layers, various
kinds of data prediction methods have been adopted. In video, the success of a 
coding method relies crucially on whether a prediction method can
truly reflect the correlation that exists between the layers. The predictive coding structure can be represented by a directed graph where a coding segment is represented as a node and an arc indicates the prediction from one coding segment to another coding segment. For bit allocation, 
we required 
the graph to have the following two properties: the
graph should be acyclic and the graph should be connected from the
source node (i.e., any node is reachable from the source node). 
The first property states that the graph has no cycle. Because a cycle can create an infinite ways to represent a coding segment for a bit-allocation, we decided to avoid such scenario. For example, a cycle of nodes A to B indicates that the coding result of A can be used to predict that of B and the result of B can then be used to predict and modify the coding result of  A. This prediction from A to B and B to A can repeat infinite times for a bit-allocation.
The second property implies that the coding object at a node
can be reconstructed based on the information on the path from the
source to that node. 

First, a DAG was formed based on a scalable coder where the basic
coding segment is a layer, and the prediction was applied on layers.
Let the number of layers of the scalable coder be $N$, denoted from
$0$ to $N-1$. We used $G = (V,A)$ to represent the DAG with node set
$V$ and arc set $A$ where the nodes correspond the layers and the arcs as the
dependency between the layers.
$G$ has a single source node (node $0$) that denotes the base layer of SC.
Arc $(i \rightarrow j)\in A$ indicates node $j$ depending on node
$i$. If we associate the (layer) node $i$ with the resolution
$i$, then the number of nodes in $G$ is the number of
resolutions. 
To reproduce the coding object at resolution $i$, we used
the required layers for the resolution and their dependency,
corresponding to the smallest connected sub-graph, denoted as $\pi_i^g$, of $G$ containing
all the paths from the node $0$
to the node $i$.  
Let $p(i)$ denote the parent nodes of node $i$ in $\pi_i^g$. The
reconstructed object at the resolution $i$ depends on the
reconstructed object at the resolutions of $p(i)$.  Figure \ref{DAG} illustrates a DAG representation of a scalable codec that supports five resolutions, where the base resolution is at node $0$.

Let us take H.264/SVC\footnote{Currently, the scalable scheme of
H.265 is inherited from H.264/SVC. \cite{Sullivan12}} as an example
\cite{SchwarzMW07}. In H.264/SVC, there are temporal prediction,
spatial prediction, and quality prediction that can remove redundancy
between the adjacent temporal layers, spatial layers, and quality
layers, respectively. The temporal prediction can exist with spatial or
quality prediction, but the spatial and quality predictions cannot be
applied to predict one layer at a time. Therefore, a temporal node
can be directed from another temporal node, and simultaneously from
either a quality or a spatial node. Depending upon the application's
environment, the coding structure, which specifies dependency
between the layers, was described in the configuration file. Figures
\ref{SVCGraph1} and \ref{SVCGraph2} show the DAG models
corresponding to two coding structures of H.264/SVC.

\section{Multi-Objective Bit-Allocation Problem} \label{secmop}

The bit-stream of SC was generated to support scalability in various dimensions. This suggests that the bit-allocation procedure
can be regarded as a multi-valued function that maps a
bit-allocation vector into a vector-valued function.

Let $G$ be the DAG constructed from the coding dependency of an SC with $N$ layers (coding segments), represented by $0$ to $N-1$, and $N$ resolutions, also represented by $0$ to $N-1$. Let 
$b$ be the bit budget and $b_i$ be the number of bits assigned to
layer $i$. Then, the bit-allocation vector $\underline{b} =
[b_i]_{i=0}^{N-1} \in R^{N}_+$ satisfies $\sum_{i=0}^{N-1} b_i \le
b$ and $b_i \ge 0$.  Let $\pi_i^g$ denote the sub-graph of $G$ for resolution $i$.  If there is more than one prediction path from resolution $0$ to resolution $i$, then $\pi_i^g$ represents the union of the paths. 
If $g$ denotes the distortion of the
reconstructed coding object against the original object $f$ and let
$E(f,G,b)$ denote the procedure of allocating $b$ bits for object $f$
with graph $G$, we have
\begin{eqnarray}
E(f, G,b ):   \underline{b}\rightarrow
[g_0(\underline{\pi}_0(\underline{b})), 
\cdots,g_{N-1}(\underline{\pi}_{N-1}(\underline{b}))]^T,\nonumber 
\label{performance}
\end{eqnarray}
where $\underline{\pi}_i(\underline{b})$ denotes the bit-allocation profile of the bit-allocation $\underline{b}$ assigned to the nodes of sub-graph $\pi_i^g$, and $g_i(\underline{\pi}_i(\underline{b}))$
measures the distortion\footnote{A main goal of SC is to
maximize the peak-signal-to-noise-ratio (PSNR) at each resolution. PSNR
is $10\log_{10} \frac{255^2}{MSE} dB,$ where $MSE$ is the
reconstruction error. 
Thus, maximizing PSNR of a resolution can be regarded as minimizing $\log MSE$ at the resolution.} of the reconstructed coding object at resolution $i$.
Then, the bit-allocation problem can be formulated as the following constrained vector-valued optimization problem:
\begin{eqnarray}
\begin{cases}
\min_{\underline{b}} \;\; [g_0(\underline{\pi}_0(\underline{b})), \cdots, g_{N-1}(\underline{\pi}_{N-1}(\underline{b}))]^T  \\
\hspace{0.5in} b_i \ge 0, \;\;\; i = 0, \cdots, N-1; \\
\hspace{0.5in}  \sum_{i=0}^{N-1} b_i \leq b, 
\end{cases} \label{bitallocate0}
\end{eqnarray}
where the bits allocated to the sub-graph $\pi_i^g$ 
are $\sum_{j \in \pi_i^g} b_j$, which is the total bits allocated to the layers that support the resolution.
We use $\Omega$ to denote the set of feasible bit-allocation
vectors of (\ref{bitallocate0}). Since $\Omega$ is the
intersection of half-spaces and hyperplanes, $\Omega$ is a convex set.

To lighten the notation, let us define the vector-valued distortion
$\underline{g}_{\Omega}(\underline{b})$ as a feasible distortion (the distortion generated by a feasible coding path in SV):
\begin{eqnarray}
\underline{g}_{\Omega}(\underline{b}) =
 [g_0(\underline{\pi}_0(\underline{b})), 
\cdots, g_{N-1}(\underline{\pi}_{N-1}(\underline{b}))]^T  \text{ when } \underline{b} \in \Omega.
\label{gfunction}
\end{eqnarray}
We also denote the feasible distortion region, the distortions derived by all the feasible coding paths, as
\begin{equation}
\underline{g}(\Omega) = \{
\underline{g}_{\Omega}(\underline{b})\}. \label{gfunction1}
\end{equation}
The optimum bit-allocation $\underline{b}^*$ can be defined as the
bit-allocation that yields the smallest distortion in each
resolution, {\em i.e}. $\underline{g}_{\Omega}(\underline{b}^*) \le \underline{g}_{\Omega}(\underline{b})$
for all $\underline{b} \in \Omega$. In other words, the optimum
bit-allocation is the minimum of the problem in
(\ref{bitallocate0}). Unfortunately, as shown in Figure
\ref{Optimumpoint}, the existence of the optimum bit-allocation
vector is uncommon. In general, we cannot compare the distortion
vectors of any two feasible bit-allocations. Two feasible  distortions can only be compared
when they are partially ordered with respect to $R_{+}^{N}$, {\em i.e}. $\underline{g}_{\Omega}(\underline{b}_1) \leq
\underline{g}_{\Omega}(\underline{b}_2)$ if and only if
$\underline{g}_{\Omega}(\underline{b}_2) -
\underline{g}_{\Omega}(\underline{b}_1) \geq \underline{0}_{N}$. By
virtue of partial ordering, there are actually many optimal
(minimal) bit-allocation solutions with respect to $R_{+}^N$ and due to this reason 
the optimum bit allocation problem for SC does not follow the
conventional assumption of the existence of the optimum
bit-allocation. Nevertheless, the optimal solutions can be derived from the study of the
multi-objection optimization problem.

The concept of optimal solutions of a multi-objective optimization
problem with respect to nonnegative orthant cone $R_{+}^N$ was
first proposed by Pareto in $1896$ \cite{Par}. Pareto defined an
optimal solution as a point in a feasible space that is
impossible to find a way of moving from, even slightly, and still
reach the consensus of all individual participants. In other words,
an optimal solution is an equilibrium position in the sense that
any small displacement in departing from the position necessarily
has the effect of increasing the values of some individual functions
while decreasing those of the other functions. In honor of Pareto,
these equilibrium positions are today called Pareto optimal points.

\subsection{Pareto Optimal Bit-Allocations}

The Pareto optimal solution deals with the case in which a set of
feasible objective vector-values does not have an optimum element. The Pareto optimal solution and the weakly Pareto optimal
solution for the bit-allocation problem are defined as follows.

The Pareto optimal bit-allocation $\underline{b}^*$ is defined as
no $\underline{b} \in
\Omega$ so that $\underline{g}_{\Omega}(\underline{b}) <
\underline{g}_{\Omega}(\underline{b}^*)$. This definition signifies  
\begin{equation}
(\underline{g}_{\Omega}(\underline{b}^*) - R_{+}^N) \cap
\underline{g}(\Omega) = \{\underline{g}_{\Omega}(\underline{b}^*)\},
\label{Pareto}
\end{equation}
where $\underline{g}_{\Omega}(\underline{b}^*) - R_{+}^N$ is the Minkowski sum\footnote{
Minkowski sum: $S + T = \{ s + t | s\in S \text{ and } t \in T\}$.}of $\underline{g}_{\Omega}(\underline{b}^*)$ and $R_{-}^N$.
The set of Pareto bit-allocations is denoted as ${\cal B}(\Omega) =
\{\underline{b}^* | \underline{b}^* \text{ satisfies }
(\ref{Pareto})\}$. In addition, the set of Pareto optimal points is
denoted as
\begin{equation}
Pareto(\underline{g}(\Omega)) =
\{\underline{g}_{\Omega}(\underline{b})|\; \underline{b} \in {\cal
B}(\Omega)\}.
\end{equation}
The bit-allocation $\underline{b}^* \in \Omega$ is called a weakly
Pareto bit-allocation if  there is no $\underline{b} \in
\Omega$ so that $\underline{g}_{\Omega}(\underline{b}) \ll
\underline{g}_{\Omega}(\underline{b}^*)$. In other words, 
\begin{equation}
(\underline{g}_{\Omega}(\underline{b}^*) - int(R_{+}^N)) \cap
\underline{g}(\Omega) = \emptyset,
\end{equation}
where $int(R_{+}^N)$ is the interior of $R_{+}^N$ and $\emptyset$ is
the empty set. 
The set of weakly Pareto
bit-allocations is denoted as ${\cal B}_w(\Omega)$ and the set of
weakly Pareto optimal points is the image of ${\cal B}_w(\Omega)$:
\begin{equation}
Pareto_w(\underline{g}(\Omega)) =
\{\underline{g}_{\Omega}(\underline{b})|\; \underline{b} \in {\cal
B}_w(\Omega)\}.
\end{equation}

A Pareto optimal bit-allocation is a weakly Pareto bit-allocation
because for a bit-allocation $\underline{b}^*$, if there is no $\underline{b}$ such that
$\underline{g}_{\Omega}(\underline{b}) <
\underline{g}_{\Omega}(\underline{b}^*)$, then, obviously, there is no $\underline{b}$ such that
$\underline{g}_{\Omega}(\underline{b}) \ll
\underline{g}_{\Omega}(\underline{b}^*)$.
Figure \ref{WeakPeratoPoint} illustrates
the Pareto optimal and weakly Perato optimal points for a
bi-criteria example.

\subsection{The Scalarization Approach} \label{secSA}

The weighted sum scalarization approach, which transforms a
vector-valued optimization problem into a scalar-valued optimization
one, is widely used to find the (weakly) Pareto optimal points
of a multi-objective optimization problem \cite{Ehr,Eic}. 
By virtue of the approach, the optimization problem in
(\ref{bitallocate0}) is transformed to solve
\begin{equation}
\min_{\underline{b}} \underline{w}^T
\underline{g}_{\Omega}(\underline{b}) = \min_{\underline{b}\in
\Omega} \sum_{i=0}^{N-1} w_i \; g_i(\underline{\pi}_i(\underline{b})), \label{scaleprob}
\end{equation}
where $\underline{w} = ([w_i]_{i=0}^{N-1})^T$ is the weight vector
with $w_i \ge 0$ for each $i$ and $\sum_{i=0}^{N-1} w_i =1$, 
$\underline{g}_{\Omega}(\underline{b})$ is a feasible distortion, and $g_i(\underline{\pi}_i(\underline{b}))$, defined in  (\ref{gfunction}), is a feasible distortion at resolution $i$. As shown in Figure \ref{Scalarization}, the optimum bit-allocation
occurs when the hyperplane tangential
to $\underline{g}(\Omega)$ has the smallest intercept  among all the parallel hyperplanes hat intercept $\underline{g}(\Omega)$.

Let $\underline{b}^*$ be the optimum bit-allocation of
(\ref{scaleprob}) with the weight vector $\underline{w}$. 
We denote that $\underline{y}(\underline{b}^*)
= ([y_i(\underline{b}^*)]_{i=0}^{N-1})^T$
satisfies the equation
\begin{equation}
\sum_{i=0}^{N-1} w_iy_i(\underline{b}^*) =  \min_{\underline{b} \in
\Omega} \sum_{i=0}^{N-1} w_i \; g_i(\underline{\pi}_i(\underline{b}))
\label{opt}
\end{equation}
and define the set of solutions of (\ref{opt}) for all normalized
weight vectors as
\begin{equation}
S_0 = \{\underline{y}(\underline{b}^*) | \text{ there is
$\underline{w} \ge \underline{0}$ with $\sum_{i}w_i = 1$ so that
$\underline{y}(\underline{b}^*)$ satisfies (\ref{opt})} \}.
\end{equation}

In general, $S_0$ is a subset of the Pareto points. As shown in Figure \ref{Scalarization}, the Pareto point $\underline{a}$ is not in $S_0$. The main result for the weighted sum scalarization
approach for solving the multi-objective optimization problem is the
equivalence of $S_0$ and the weakly Parent optimal points when
$\underline{g}(\Omega) + R_{+}^N$
is a convex set. Figure \ref{peter} illustrates an example where $\underline{g}(\Omega)$ is not convex, but $\underline{g}(\Omega) + R_{+}^N$ is a convex set. The result is stated through the following theorem.

\noindent{\textbf{Theorem 1} \cite{Ehr}. If $\underline{g}(\Omega)
+ R^N_{+}$ is a convex set, then $S_0 =
Pareto_w(\underline{g}(\Omega))$.

\noindent The theorem indicates that if $\underline{g}(\Omega) +
R^N_{+}$ is a convex set, then the scalarization approach can
determine nothing but all weakly Pareto points and weakly Pareto
bit-allocations of $\underline{g}(\Omega)$.

\section{Main Results} 
\label{RDModel}

Since it is important and insightful to have all alternatives available for decision makers to choose which Pareto point to operate on, the primary purpose here is to derive a sufficient condition so that $\underline{g}(\Omega) + R_{+}^N$ is a convex set. 


The distortion space at a resolution is defined as all the feasible distortions that the resolution can generate from a given bit budget. As shown in Figure \ref{rd}, if the bit budget is $b$, then the distortion at resolution $i$ is defined as the set, 
\begin{equation}
\{g_i(\underline{\pi}_i(\underline{b})) | g_i(\underline{\pi}_i(\underline{b})) \text{ is the $i$-th component of a } \underline{g}_{\Omega}(\underline{b}) \in \underline{g}(\Omega)\},
\end{equation} where $\underline{g}_{\Omega}(\underline{b})$ and  $\underline{g}(\Omega)$ are defined in (\ref{gfunction}}) and (\ref{gfunction1}), respectively, and $\underline{\pi}_i(\underline{b})$ is defined in (\ref{performance}) as the bit-allocation profile of resolution $i$ in the DAG. Hereafter, let bit-rate $r_i$ denote the total number of bits in the bit profile $\underline{\pi}_i(\underline{b})$ assigned to the resolution $i$ in the DAG. Note that many bit-allocation profiles assign the same total number of bits $r_i$ at resolution $i$.
Let $D_i(r_i)$ denote the rate-distortion (R-D) function of $r_i$ at resolution $i$.
The R-D function is the lower envelope formed by all the distortions at the resolution $i$ that can be obtained by coding an image or a video with bit-rate $r_i$. 


The main result is summarized in Theorem 2, which indicates that  the convex set $\underline{g}(\Omega) + R_{+}^N$ can be characterized from the R-D function of each resolution of a scalable
coder.  The Lemma 1 indicates that $\underline{g}(\Omega) + R_{+}^N$ is equivalent to  $Pareto_w(\underline{g}(\Omega)) + R_{+}^N$.

\noindent{\textbf{Lemma 1}}. 
 \begin{eqnarray}
\underline{g}(\Omega) + R_{+}^N  & = &
Pareto_w(\underline{g}(\Omega)) + R_{+}^N. \label{wgeq}
\end{eqnarray}
\noindent{\textbf{\emph{Proof}}: } \\
Clearly,  $\underline{g}(\Omega) + R_{+}^N  \supseteq
Pareto_w(\underline{g}(\Omega)) + R_{+}^N$, as $Pareto_w(\underline{g}(\Omega))$ is a subset of $\underline{g}(\Omega)$. To show the other direction:  let $\underline{d}$ be a point in $\underline{g}(\Omega) + R_{+}^N$ and  the bit-allocation of $\underline{d}$ is $\underline{b} = [b_i]_{i=0}^{N-1}$. Then, it is clear that $\sum_{i=0}^{N-1} b_i  \le  b$.  Since there is a weakly Pareto point $\underline{d}^w$ with bit-allocation $\underline{b}^w = [b_i^w]_{i=0}^{N-1}$ with $\sum_{i=0}^{N-1} b_i^w = b$ such that $\underline{d}^w \le \underline{d}$. Therefore, $Pareto_w(\underline{g}(\Omega)) + R_{+}^N \supseteq  \underline{g}(\Omega) + R_{+}^N $. 

\noindent \textbf{\emph{End of Proof}}. 

Under mild assumptions on the distortion space $\underline{g}(\Omega)$ and R-D functions, the below lemma indicates that any feasible distortion $\underline{g}_{\Omega}(\underline{b})$ can be represented by the R-D functions.

\noindent{\textbf{Lemma 2}}. 
Let the feasible distortion space $\underline{g}(\Omega)$ be a compact region and let $D_i(r_i)$ be the R-D curve of resolution $i$ with  
$r_i \le b$. 
If $\{D_i(r_i)\}$ are strictly decreasing convex functions, then there are one-to-one and onto functions $\{q_i\}$ that map the $i$-th component $g_i(\underline{\pi}_i(\underline{b}))$ of any feasible distortion to the bit-rate  in $[0,b]$ so that  
\begin{equation}
g_i(\underline{\pi}_i(\underline{b})) = D_i(q_i(g_i(\underline{\pi}_i(\underline{b})))) \text{ for all resolution $i$ and all feasible bit-allocations $\underline{b}$}. \label{qfunction}
\end{equation}
Meanwhile, $q_i$ is a strictly concave function. \\
\noindent{\textbf{\emph{Proof}}: } \\
Without loss of any generality, we can use a two-resolution example to sketch the main concept of the proof. Figure \ref{mapping} illustrates the example where the minimum and the maximum distortions with bit budget $b$ for resolution $1$ are $A$ and $B$, respectively. The $q_1$ is a one-to-one and onto mapping of the vertical segment $[B,A]$ at $b$ in the right sub-graph to the bit-rates $[0,b]$. The horizontal dashed line in the left sub-figure shows the distortion of resolution $1$ varies with a fixed distortion of resolution $2$. The dashed line intersects the distortion space at an interval with end points at $C$ and $D$. Since $D_1$ is a strictly decreasing convex function, as shown in the right sub-figure, the interval $[C, D]$ has a unique corresponding curve in $D_1$ and the domain of the curve is defined from $q_1(D)$ to $q_1(C)$.  On the other hand, similar discussions can imply that the mapping $q_2$ is an one-to-one and onto mapping of the distortion at resolution $2$ to the bit-rates in the domain of the R-D function $D_2$. This concludes that any distortion point in $\underline{g}(\Omega)$ can be represented based on the R-D functions and the mapping $q_1$ and $q_2$.  Since $q_1$ is the inverse function of the strictly convex function $D_1$, $q_1$ is a strictly concave function \cite{Boy}. This can also be observed at the right sub-figure of Figure \ref{mapping} that the function $q_1$ maps intervals $[A,B]$ to $[b, 0]$.
The mathematical induction can then be used to extend the proof for cases with more than two resolutions. 

\noindent \textbf{\emph{End of Proof}}.\\
The following lemma indicates that if the weakly Pareto points are continuous, and $\underline{a}$ and $\underline{b}$ are two weakly Pareto points, then any weakly Pareto points from $\underline{a}$ to $\underline{b}$ must be located either inside or within the axis-aligned (minimum) bounding box of $\underline{a}$ and $\underline{b}$\footnote{The axis-aligned minimum bounding box for a given point set is its minimum enclosing box subject to the constraint that the edges of the box are parallel to the coordinate axes.} .

\noindent{\textbf{Lemma 3}}. If the weakly Pareto point form a continuous curve (surface) and $[a_i]_{i=0}^{N-1}$ and $[b_i]_{i=0}^{N-1}$ be any two weakly Pareto points, then any weakly Pareto point from $[a_i]_{i=0}^{N-1}$ to $[b_i]_{i=0}^{N-1}$ should be either inside or in the axis-aligned minimum bounding box of $[a_i]$ and $[b_i]$ and can be represented as $[p_i(t)]_{i=0}^{N-1}$, where $t \in [0,1]$, and 
\begin{equation}
p_i(t) = a_i + \alpha_i(t) (b_i - a_i) = (1 - \alpha_i(t)) a_i + \alpha_i(t) b_i, \label{param}
\end{equation}
where $\alpha(t)$ is a continuous, $\alpha_i(t) \in [0, 1]$, and $\alpha(0) = 0$ and $\alpha(1) = 1$.

\noindent{\textbf{\emph{Proof}}: } \\
We will prove this lemma by using mathematical induction on the dimension of the distortion space $\underline{g}(\Omega)$ with coordinate axes $[g_0, \cdots, g_{N-1}]$.  For a two-dimensional distortion space,  let $[a_0, a_1]$ and $[b_0, b_1]$ be two weakly Pareto points and let $B^2$ denote the axis-aligned minimum bounding box of $[a_0, a_1]$ and $[b_0, b_1]$. Since the weakly Pareto points between $[a_0, a_1]$ and $[b_0, b_1]$ are continuous, if there is a $[c_0, c_1]$ inside $B^2$ such that either the horizontal line, $g_1 = c_1$, or the vertical line, $g_0 = c_0$, intersects the continuous Pareto curve at a point $[d_0, d_1]$ that is outside $B^2$, then one of the weakly Pareto points $[d_0, d_1]$, $[a_0, a_1]$, and $[b_0, b_1]$ would not be a weakly Pareto point, depending on the location of the intersection point as shown in Figure \ref{boundingbox}. Therefore, all the weakly Pareto points between $[a_0, a_1]$ and $[b_0, b_1]$ must be inside or in $B^2$ and, hence, can be represented as (\ref{param}).

Let us assume that the lemma is true up to dimension $n-1$. Let $[a_i]_{i=0}^{n-1}$ be $[b_i]_{i=0}^{n-1}$ be two weakly Pareto points in an $n$-dimensional distortion space with coordinates $[g_0, \cdots, g_{n-1}]$, and let $B^n$ be the axis-aligned minimum bounding box  of $[a_i]_{i=0}^{n-1}$ and $[b_i]_{i=0}^{n-1}$. Then, for any point $[c_i]_{i=0}^{n-1}$ inside $B$, there are $n$ axis-aligned hyperplanes, $g_0= c_0$, $\cdots$, $g_{n-1} = c_{n-1}$. Without loss of any generality, let us take the hyperplane $g_{n-1} = c_{n-1}$. This hyperplane intersects the continuous Pareto curve in a $(n-1)$-dimensional axis-aligned minimum bounding box $B^{n-1}$ of $[a_0,\cdots, a_{n-2}, c_{n-1}]$ and $[b_0,\cdots, b_{n-2}, c_{n-1}]$. Let $[d_0, \cdots, d_{n-2}, c_{n-1}]$ be an intersection point, then by mathematical induction, $[d_0, \cdots, d_{n-2}, c_{n-1}]$ must be inside or in the bounding box $B^{n-1}$. As a result, the point $[d_0, \cdots, d_{n-2}, c_{n-1}]$ is also inside or in the bounding box $B^n$. Since $[c_i]_{i=0}^{n-1}$ is any point inside $B^n$, we conclude that the lemma in true for dimension $n$.

\noindent \textbf{\emph{End of Proof}}.\\

\noindent{\textbf{Theorem 2}}. 
Let the feasible region $\underline{g}(\Omega)$ be a compact region, $D_i(r_i)$ be the R-D function of resolution $i$ with $r_i \le b$,  and $q_i$ be the mapping derived in Lemma 2. If $\{D_i(r_i)\}$ are strictly decreasing convex functions and if the weakly Pareto points of $\underline{g}(\Omega)$ forms a continuous curve, then $\underline{g}(\Omega) + R_{+}^N$ is a convex set. 

\noindent{\textbf{\emph{Proof}}:} \\
By Lemmas 1 and 2, for any two points in $\underline{g}(\Omega) + R_{+}^N$, $[u_i]_{i=0}^{N-1}$ and $[v_i]_{i=0}^{N-1}$, we can find two weakly Pareto points $\underline{D}^0 =  [D_i(r_i^0)]_{i=0}^{N-1}$ and $\underline{D}^1 = [D_i(r_i^1)]_{i=0}^{N-1}$  with $b \ge r_i^0 \ge q_i(u_i)$ and $b \ge r_i^1 \ge q_i(v_i)$ such that 
\begin{equation}
\underline{D}^0 \le [u_i]_{i=0}^{N-1} \text{  and } \underline{D}^1 \le [v_i]_{i=0}^{N-1}. \label{eqn}
\end{equation}
 
To simplify the notation, we let $a_i = D_i(r_i^0)$ and $b_i = D_i(r_i^1)$. The continuous functions $\{\alpha_i(t)\}$ have the domain $t \in [0,1]$ and the range $\alpha_i(t) \in [0,1]$ and the end points $\alpha_i(0)= a_i$ and $\alpha_i(1) = b_i$. Since the weakly Pareto points form a continuous curve, according to Lemma 3, any weakly Pareto point $[p_i(t)]_{i=0}^{N-1}$ between the Pareto point $\underline{D}^0$ and $\underline{D}^1$ can be represented using $\{\alpha_i(t) \}$ as 
\begin{equation}
p_i(t) = a_i + \alpha_i(t) (b_i - a_i) = (1 - \alpha_i(t)) a_i + \alpha_i(t) b_i.
\end{equation}
As $t$ varies from $0$ to $1$, $p_i(t)$ varies continuously from $a_i$ to $b_i$.
By Lemma 2, we have
\begin{equation}
(1 - \alpha_i(t)) a_i + \alpha_i(t) b_i = D_i(q_i((1 - \alpha_i(t)) a_i + \alpha_i(t) b_i )).
\end{equation}
Since $D_i$ is a decreasing and convex and $q_i$ is concave, $D_i(q_i)$ is a convex function \cite{Boy}. Therefore, \begin{eqnarray}
D_i(q_i((1 - \alpha_i(t)) a_i + \alpha_i(t) b_i )) & \le & (1 - \alpha_i(t))  D_i(q_i(a_i)) +  \alpha_i(t)D_i(q_i(b_i)) \\
& = & (1 - \alpha_i(t))  a_i +  \alpha_i(t) b_i, \label{convex}
\end{eqnarray}
where the inequality and equality are derived from the definition of convex function and Lemma 2, respectively.
Since $a_i = D_i(r_i^0)\le u_i$ and $b_i = D_i(r_i^{1}) \le v_i$ , from Equations (\ref{eqn}) and (\ref{convex}), we have 
\begin{equation}
D_i(q_i((1 - \alpha_i(t)) a_i + \alpha_i(t) b_i ))  \le \alpha_i(t) a_i  + (1 - \alpha_i(t) ) b_i \le \alpha_i(t) u_i +
(1-\alpha_i(t)) v_i. \label{distor1}
\end{equation}

Since $[D_i(q_i((1 - \alpha_i(t)) a_i+ \alpha_i(t) b_i ))]_{i=0}^{N-1}$ for $t \in [0,1]$ are weakly Pareto points of $\underline{g}(\Omega)$,  Equation (\ref{distor1}) implies that the points lie within the line segment connecting $[u_i]_{i=0}^{N-1}$ and $[v_i]_{i=0}^{N-1}$ are in $\underline{g}(\Omega) + R_{+}^N$.
Since $\underline{u} =[u_i]_{i=0}^{N-1}$ and $\underline{v} = [v_i]_{i=0}^{N-1}$ are any two points in $\underline{g}(\Omega) + R_+^N$, we can conclude that $\underline{g}(\Omega) + R_+^N$ is a convex set.

\noindent \textbf{\emph{End of the proof}}.\\
Figure \ref{mainresult} illustrates a two-resolution example of the above theorem.
Theorem 2 provides a sufficient condition to characterize all the weakly Pareto points by using the weighted sum scalarization approach from the R-D curve of each resolution and the distortion space.
Therefore, according to Theorem
1, by using the weighted sum scalarization approach, all
weakly Pareto optimal points can be derived. 

In practice, the feasible bit-allocation space $\Omega$ and the feasible distortion space $\underline{g}(\Omega)$ of SC are discrete. Since $r_i$ are discrete, $\tilde D_i(r_i)$, called the continuous extension of $D_i(r_i)$, can be defined as a continuous function of $r_i$ which contains $D_i(r_i)$ with $r_i \in Z_{+}$ and $r_i \in [0, b]$. 
Meanwhile, the distortion $\underline{\tilde g}(\Omega)$, called the continuous extension of discrete point set  $\underline{g}(\Omega)$,  can be defined as a compact set which contains $\underline{g}(\Omega)$ so  that all weakly Pareto points of $\underline{g}(\Omega)$ are also weakly Pareto points of $\underline{\tilde g}(\Omega)$.  The following corollary is the discrete version of Theorem 2.


\noindent{\textbf{Corollary 1}}. 
Let $\tilde D_i(r_i)$ and $\underline{\tilde g}(\Omega)$ be the continuous extension of discrete function $D_i(r_i)$ and discrete ponint set $\underline{g}(\Omega)$, respectively. If $\{\tilde D_i(r_i)\}$ are strictly decreasing convex functions and if all the weakly Pareto points of $\underline{\tilde g}(\Omega)$ forms a continuous curve (surface), then all weakly Pareto points of $\underline{g}(\Omega)$ can be derived using the weighed sum scalarization approach.

\noindent{\textbf{\emph{Proof}}:} \\
According to Theorem 2, $\underline{\tilde g}(\Omega) + R_+^N$ is a convex set. Therefore, all weakly Pareto point of $\underline{\tilde g}(\Omega)$ can be derived by the scalarization approach. Since the weakly Pareto point of $\underline{g}(\Omega)$ is a subset of that of $\underline{\tilde g}(\Omega)$, using the scalarization approach, all weakly Pareto points of $\underline{g}(\Omega)$ can be derived.

\noindent \textbf{\emph{End of the proof}}.\\


\section{Conclusions}
\label{con}

To conclude, we represented the prediction structure that removes the redundancy in
scalable coding (SC) as a directed acyclic graph and formulated the optimal bit-allocation problem on the graph as a
multi-criteria optimal problem. 
In general, there can be many optimal
solutions (called Pareto points), but the
performance of those solutions are incomparable. In SC, the weighed sum scalarization approach is a popular way to derive Pareto points. Since the Pareto points derived via the weighted sum scalarization approach is a subset of all Pareto points, it is important to present the conditions in SC so that all the Pareto points can be derived through the scalarization approach. Our main results showed that if the rate-distortion (R-D) function of each resolution of a SC method  is strictly decreasing and convex and the weakly Pareto points form a continuous curve, then all the Pareto optimal solutions can be derived through the scalarization approach.  \\ 


\noindent{\bf{Acknowledgement}}: Wen-Liang Hwang would like to express his gratitude to Mr. Jinn Ho, Mr. Chia-Chen Lee, and Dr.  Guan-Ju Peng. Without their assistances, this paper cannot be finished.

\newpage
\input{ref}

\newpage
\input{figure}

\end{document}

%% file: ref.tex

%% file: figure.tex

\begin{figure}[t]
\begin{center}
\begin{tabular}{c}
\includegraphics[scale=0.4]{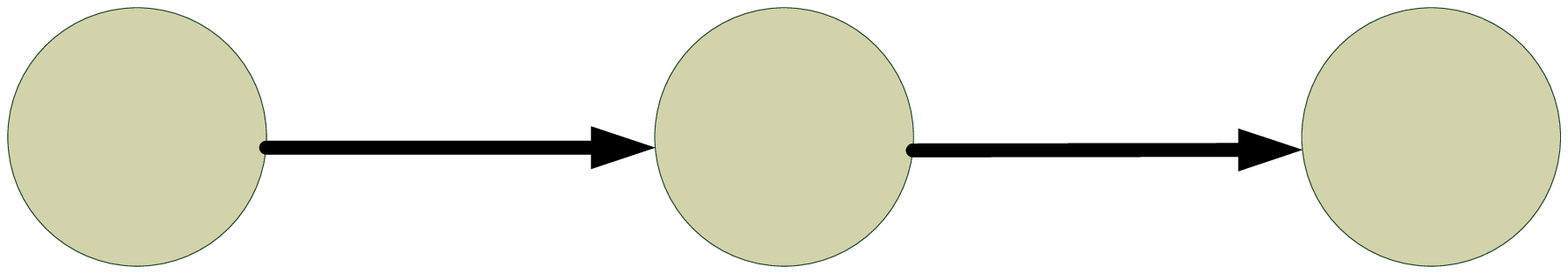} \vspace{-2in} \\
\includegraphics[scale=0.4]{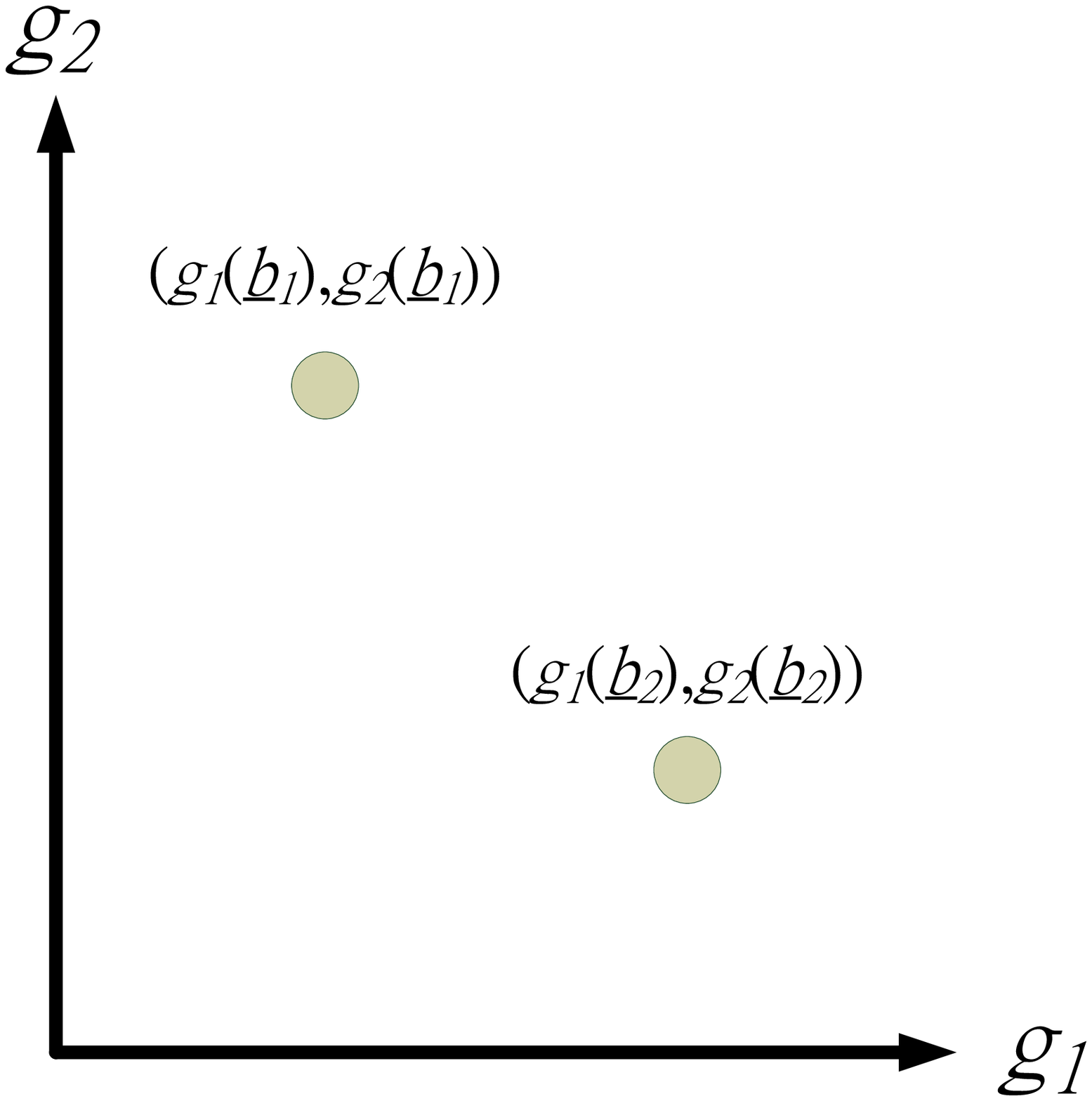} \vspace{-1in}
\end{tabular}
\caption{Top: An SC supports three resolutions: QCIF (the leftmost
node), CIF (the leftmost and the middle nodes), and HD (all three
nodes). The arcs indicate the coding dependence of resolutions. 
Bottom: a possible distortion distribution for two bit-allocations $\underline{b}_1$ and $\underline{b}_2$. Both bit-allocations assign the same bits to
the QCIF node, but assign different bits to CIF and HD nodes with $\underline{b}_1 = [b, 0]^T$ ($b$ bits are assigned to the middle node to support CIF) and $\underline{b}_2 =
[0, b]^T$ ($b$ bits are to the rightmost node to support HD). The 
distortion vectors for CIF and HD are
$[g_1(\underline{b}_1), g_2(\underline{b}_1)]^T$ and
$[g_1(\underline{b}_2),
g_2(\underline{b}_2)]^T$ for $\underline{b}_1$ and
$\underline{b}_2$, respectively. Obviously, $g_1(\underline{b}_1) <
g_1(\underline{b}_2)$ because $\underline{b}_1$ uses more bits to
encode CIF. However, we cannot determine if either
$g_2(\underline{b}_1)$ or $g_2(\underline{b}_2)$ is smaller because
$\underline{b}_1$ and $\underline{b}_2$ use the same number of bits
to encode HD. The result depends on coding algorithms and video. If
the distortions are as shown, which bit-allocation is better cannot
be determined because $\underline{b}_1$ is better for CIF, but
$\underline{b}_2$ is better for HD.}
\label{Uncomparable}
\end{center}
\end{figure}

\begin{figure}[htbp]
\centerline{\includegraphics[scale=0.6]{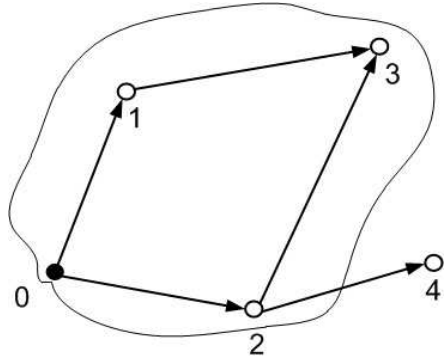}}
\caption{The DAG representation of a scalable codec that supports five resolutions. The node $0$ represents the base layer and the resolution $0$. The high-lighted sub-graph of resolution $3$, denoted as $\pi_3^g$, contains nodes $0$, $1$, and $2$. } \label{DAG}
\end{figure}

\begin{figure}
\centering \centerline{ \hfil
{\includegraphics[scale=0.4]{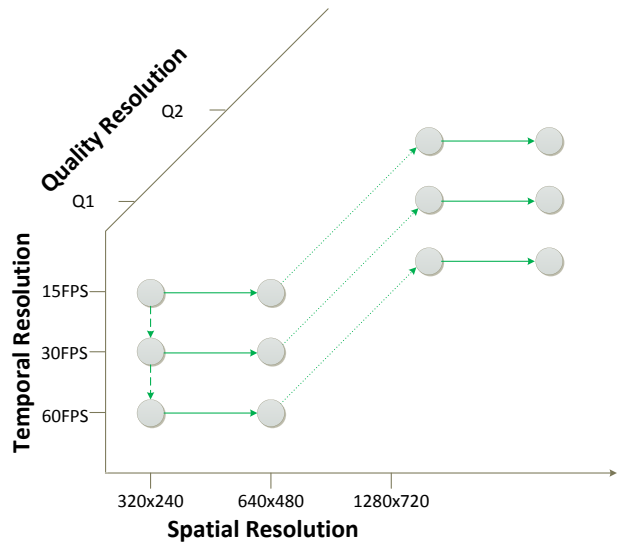}}}
 \caption{A configuration example of H.264/SVC. The DAG has twelve
 nodes (resolutions). 
 The dashed arcs (from $15$ FPS to $30$ FPS and $30$ FPS to $60$ FPS) correspond to the temporal dependency in the lowest spatial layer.
 The arcs between the spatial nodes and quality nodes are based on an inter-layer prediction technique adopted
by H.264/SVC. The base layer is at the node
$(320\times240,15\text{ FPS},\text{Q1}$). Any node is reachable from
the base layer node, and the nodes present in the path are used to
support the video resolution associated to the end node. FPS is the abbreviation for frames per second.}
\label{SVCGraph1}
\end{figure}

\begin{figure}
\centering \centerline{ \hfil
{\includegraphics[scale=0.4]{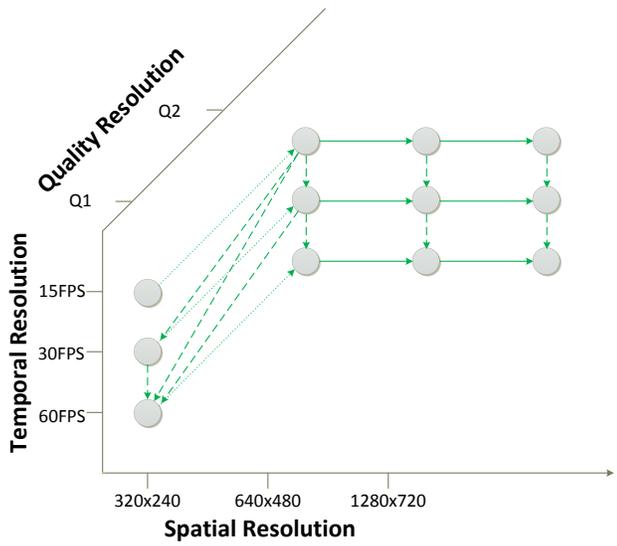}}}
\caption{ A configuration example of H.264/SVC. The DAG has 12
nodes (resolutions). The dashed arcs corresponding to temporal
dependency have a complicated ``key frame" structure between the
first two quality resolutions for the lowest spatial resolution. The
key frame technique uses the reconstructed frames at higher
spatial/quality and lower temporal resolution as a reference to
predict the frames at lower spatial/quality and higher temporal
resolution \cite{SchwarzMW07}. The temporal prediction is also
available for higher quality layers. The prediction between spatial
nodes and quality nodes is applied at the same temporal resolution.
The base layer is at $(320\times240,15\text{ FPS},\text{Q1}$). Any
node is reachable from the base layer node. Note that the node
$(320\times240,60\text{ FPS},\text{Q1}$) can be reached by more than
one path. All the nodes in the paths are used to support the video
resolution associated with the node
$(320\times240,60\text{ FPS},\text{Q1}$). } \label{SVCGraph2}
\end{figure}


\begin{figure}[p]
\centering \centerline{ \hfil
{\includegraphics[scale=0.5]{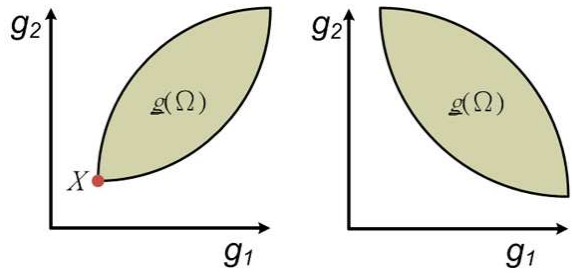}}}
\caption{Even for two resolutions, the existence of
the optimum solution is uncommon for SVC. Left: $X$ is the minimum
point because the distortion at the point in each resolution is the
smallest when compared to other points in $\underline{g}(\Omega)$. In this
case, $X$ is the optimum point. Right: This example demonstrates that the
optimum point does not always exist. In this case, there is no
optimum point.} \label{Optimumpoint}
\end{figure}

\begin{figure}
\centering \centerline{ \hfil
{
\includegraphics[scale=0.35,keepaspectratio=true, bb=0 0
650 750]{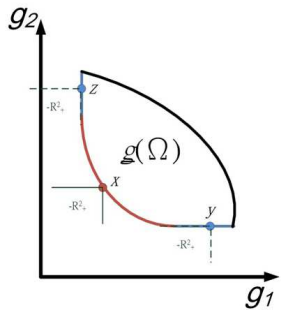}
}
}
\caption{The Pareto and weakly Pareto optimal points
for the bi-criteria case. The Pareto optimal points are the points
on the red curve, which is derived based on the closed cone $R_{+}^2$, as
shown under $X$. The weakly Pareto optimal points are the points on
the blue and red curves, which are derived based on the open cone
$(int(R_{+}^2) \cup \{0\})$, as shown under $Y$ and $Z$. }
\label{WeakPeratoPoint}
\end{figure}


\begin{figure}
\centering \centerline{ 
{\includegraphics[scale=0.45]{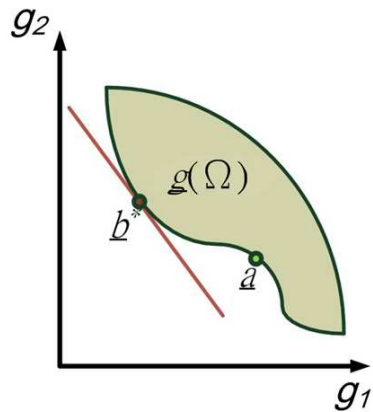}}}
 \caption{The hyperplane denoted by the red line was
determined by the weight factor $\underline{w}$. The hyperplane is
tangential to $\underline{g}(\Omega)$ at the Pareto bit-allocation
$\underline{b}^*$. The scalarization approach cannot find all
optimal bit-allocations of $\underline{g}(\Omega)$ (since
$\underline{g}(\Omega) + R_+^2$ is not a convex set). For example,
the Pareto bit-allocation $\underline{b}^*$ could be obtained, but not the
Pareto bit-allocation $\underline{a}$. }
\label{Scalarization}
\end{figure}

%
%

\begin{figure}[p]
\centering \centerline{ 
{\includegraphics[scale=0.6]{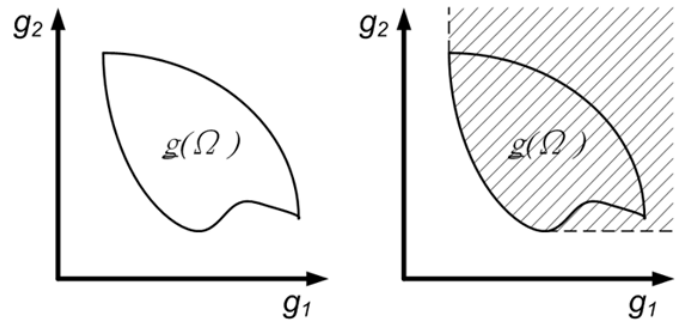}}}
\caption{An example that illustrates $\underline{g}(\Omega) + R_+^2$ is convex but $\underline{g}(\Omega)$ is not. } \label{peter}
\end{figure}

\begin{figure}[p]
\centering \centerline{ 
{\includegraphics[scale=0.55]{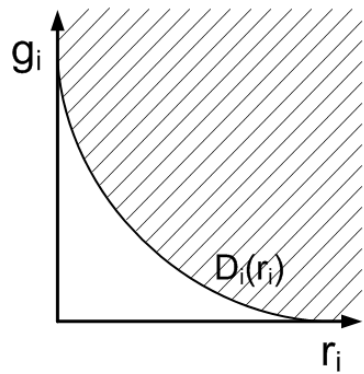}}}
\caption{Distortion space and the R-D function at a resolution. The dashed area is the distortion space, corresponding to all the feasible distortions at that resolution for a bit budget. The lower envelope of the area is the R-D curve at that resolution. } \label{rd}
\end{figure}

\begin{figure}[p]
\centering \centerline{ \hfil
{\includegraphics[scale=0.45]{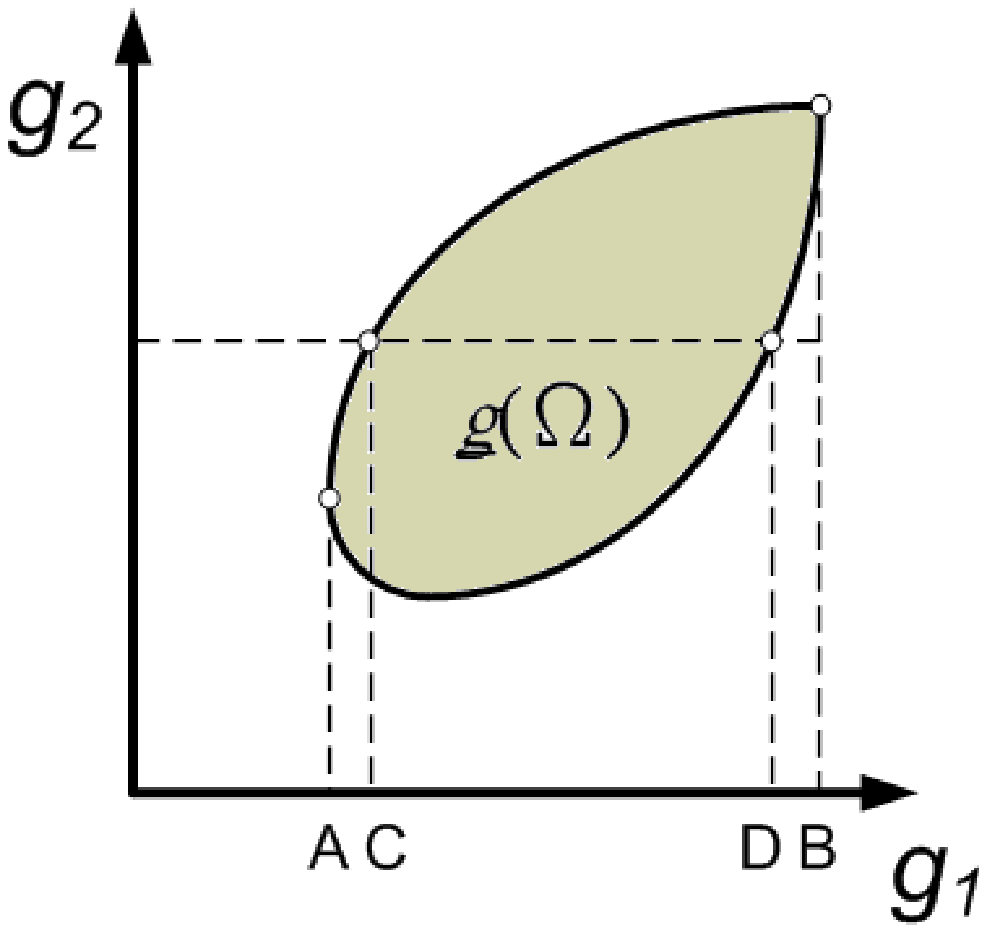}}
{\includegraphics[scale=0.45]{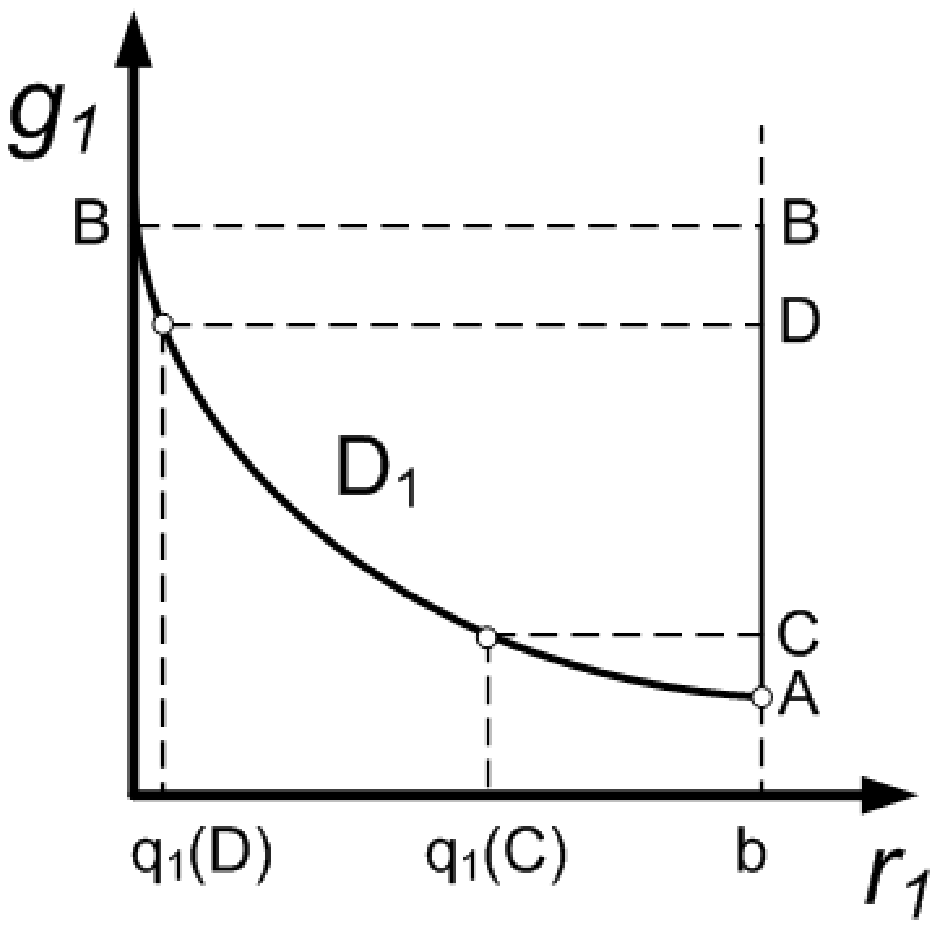}}}
\caption{Left: the dashed line indicates that the distortion for resolution $2$ is fixed and the distortion for resolution $1$ varies from $C$ to $D$. $A$ is the minimum distortion of resolution $1$ with bit budge $b$ and $B$ is the maximum distortion of the resolution. Right: each point in the vertical segment $[A, B]$ has a unique corresponding point in $D_1$ because $D_1$ is a strictly non-increasing convex function. This example illustrates that $q_1$ is one-to-one and onto function from distortion to bit-rate.  } \label{mapping}
\end{figure}

\begin{figure}[!htbp]
\centerline{\includegraphics[scale=0.45]{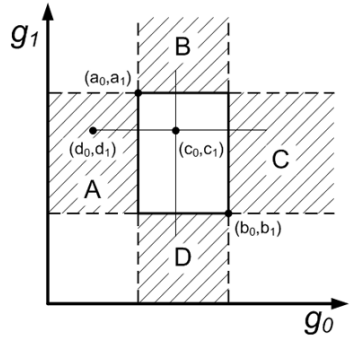}}
\caption{$[c_0, c_1]$ is any point inside the axis-aligned minimum bounding box of $[a_0,a_1]$ and $[b_0, b_1]$. $[d_0,d_1]$ is an intersection point of the curve of weakly Pareto points from one weakly Pareto point $[a_0,a_1]$ to the other weakly Pareto point $[b_0, b_1]$ and the horizontal line $g_1 = c_1$. Therefore, $d_1 = c_1$. Since $d_0 < a_0$ and $c_1 < a_1$, $[a_0, a_1]$ is not a weakly Pareto point.  Similarly, if the intersection point is in the regions B or C, the point is not a weakly Pareto point and if the intersection point is in region D, then $[b_0, b_1]$ is not a weakly Pareto point. }
\label{boundingbox}
\end{figure}

\begin{figure}[!htbp]
\centerline{\includegraphics[scale=0.6]{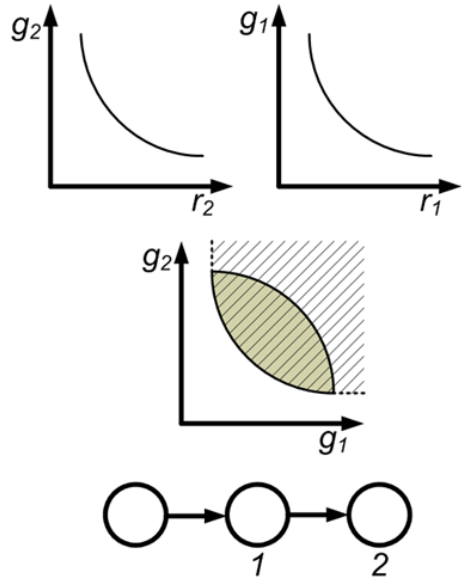}}
\caption{Characterization of the Pareto points from the R-D functions of resolutions 1 and 2: if R-Ds are strictly non-increasing convex functions, as shown in the top left and top right subfigures, then $\underline{g}(\Omega) + R_{+}^2 $ is also convex, as shown in the bottom subfigure.}
\label{mainresult}
\end{figure}